# Visual Fixations Duration as an Indicator of Skill Level in eSports

Boris B. Velichkovsky[1], Nikita Khromov[2], Alexander Korotin[2], Evgeny Burnaev[2], Andrey Somov[2]

[1]Moscow State University, Mokhovaya 11/5, 125009 Moscow, Russia
`velitchk@mail.ru`
[2]Skolkovo Institute of Science and Technology, Bolshoy Boulevard 30, bld. 1, 121205 Moscow, Russia
`{nikita.khromov, a.korotin, e.burnaev, a.somov}@skoltech.ru`

**Abstract**. Using highly interactive systems like computer games requires a lot of visual activity and eye movements. Eye movements are best characterized by visual fixation – periods of time when the eyes stay relatively still over an object. We analyzed the distributions of fixation duration of professional athletes, amateur and newbie players. We show that the analysis of fixation durations can be used to deduce the skill level in computer game players. Highly skilled gaming performance is characterized by more variability in fixation durations and by bimodal fixation duration distributions suggesting the presence of two fixation types in high skill gamers. These fixation types were identified as ambient (automatic spatial processing) and focal (conscious visual processing). The analysis of computer gamers' skill level via the analysis of fixation durations may be used in developing adaptive interfaces and in interface design.

## 1    Introduction

eSports involves highly dynamic interactions with complex computer-mediated environments. It is a complex activity the success of which depends on many factors – technological, cognitive, personal, and even social. From the Human-Computer Interaction (HCI) perspective, eSports is an interesting activity as it provides for the study of human-user interaction under the conditions of stress and the need for high performance [8,10]. For instance, it is vital for the study of the visual aspects of the interaction as contemporary user interfaces are still mainly visually oriented. This suggests the importance of studying the patterns of eye activity in eSports as the eye movements are the most important prerequisite for the gathering of visual information [11].

Human vision is an active vision system [3]. It depends on the eye movements to allow for a detailed exploration of objects in the environment. That is why the analysis of the eye movements can reveal a lot about the organization of visual information processing. Today, the analysis of eye movements is made much easier by the availability of affordable non-invasive video-based eye-trackers. For instance, eye movements are often studied in users interacting with computers [1,9].

Eye movements are complex and have many subtypes. Important events within the eye activity are fixations – time intervals during which the eyes stand relatively still (although there is always some jitter) with the gaze pointed at an external object under detailed inspection [13]. Visual perception is said to be accomplished during fixations while there is no visual processing during the ballistic eye movements between fixations (saccades, "saccadic suppression"). Fixations are characterized by (1) their location within the visual field and (2) by their duration. It is the fixation duration that we focus on in this paper. Fixation durations differ a lot during eye movements. Typical durations may be as short as 50 ms and as long as over 2000 ms. The distributions of fixation durations are skewed to the left with typical median durations 0f 200-250 ms, mean durations of 300-350 ms, and a long right tail of long and very long fixations. Fixation durations are used for a distinction between subtypes of fixations with "short" fixations considered ambient (spatial processing) and "long" (over 300-500 ms) fixations considered focal (conscious perception) [5, 12]. Fixation durations are sensitive to many factors like the skill level of the user and the cognitive load experienced by the user.

In this paper, we study the differences in fixation durations between eSports athletes of different proficiency levels (amateurs with less experience, experienced amateurs, and professional players) during the gameplay. To this end, we computed the individual distributions of fixation durations and compared various parameters characterizing the form of these distributions. The rationale was to discover how differences in the parameters of individual fixation durations distributions may reflect differences in the eSports athletes' proficiency level under realistic conditions of an emotion-provoking competition (Counter-Strike: Global Offensive, CS:GO).This study is novel to HCI in that it, for the first time, uses the analysis of individual fixation duration distributions to measure skill level to drive possible human-computer interface changes in a high-performance HCI scenario.

## 2 Method

### 2.1 Subjects

In this experiment players with different CS:GO experience were invited. Some players have had only limited CS:GO experience (amateurs with less than 700 hours playing the game). This was the low skill group (N=10). Other amateur players have had more experience with the game (over 700 hours) and were in the high skill group (N=7). There also were some professional eSports athletes (N=4) with over 10.000 hours of playing the game which means they play more than 4 hours per day (they are designated as the pro group below) The athletes are from the professional Monolith team, Russia. This team is affiliated with the Skoltech Cyberacademy. Amateur play-ers are Skoltech MS/PhD students as well as research and administrative staff from Skoltech.

### 2.2 Equipment

An in-house IoT platform for the eSports data collection and analysis was used. The platform block diagram is shown in Figure 1. It includes three units: a sensing unit, a gaming PC, and a CounterStrike (CS:GO) server. It allows for data collection from heterogeneous sensors and a game service deployed on the CS:GO server.

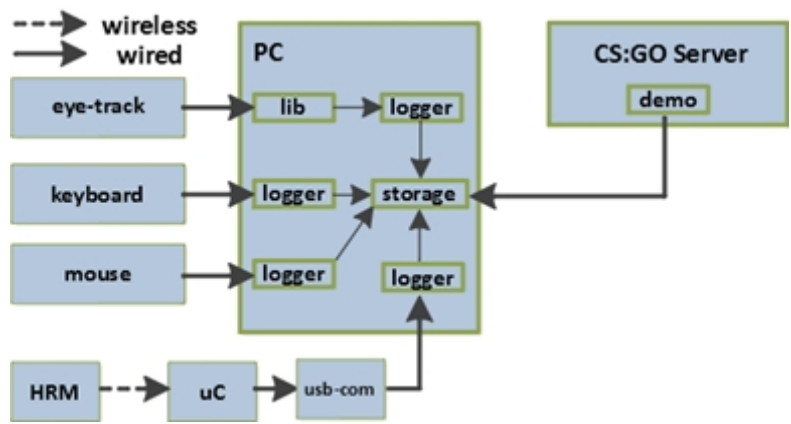

**Fig. 1.** Block diagram for the platform for eSports data collection

In terms of sensing, the proposed platform contains the following sensors: Tobii EyeX eye-tracker, Garmin Heart Rate Monitor (HRM) belt, key/mouse logger, in-game data logger (CSGO HLTV demo). We used the Tobii eye-tracking library for eye-position measurements and developed a custom script for capturing the data-stream from the Tobii library and recording these data to a local .csv text file. We measured the gaze position at 30 Hz. PC unit is an advanced (storage and processing capability) gaming PC to which all external physiological sensors, i.e. eye-tracker and HRM, are connected. Apart from the physiological data we performed logging of mouse and keyboard on the PC, as well as collection of game statistics

### 2.3 Game Scenario

There are two roles for the Retake modification of CS:GO discipline: a terrorist team and counter-terrorist team. The terrorist team (2 players) is characterized by the defensive role. They have a bomb planted on the territory and have to defend it. The counter-terrorist team (3 players) has to deactivate the bomb or to kill the enemy team. The system shows the bomb location on the map at the beginning of each round. Players are also asked to buy exactly the same set of weapons for each round. Each round lasts for 40 seconds while there are 12 rounds in total. This scenario must be played without any breaks between the rounds.

### 2.4 Procedure

Prior to participate in the experiment, all participants were informed about the project, its goals, and the experiment. We obtained a written consent from each participant in the experiment. Afterwards, we ensured by a questionnaire that all the participants are in a good form and do not take any drugs in order to avoid the interference with the experimental results

### 2.5 Fixation Extraction

Raw eyegaze data (x- and y-coordinates) were obtained from the eyetracker. Fixations were extracted in the R statistical computing environment using the *gazepath* package [11]. For fixation extraction, the *Mould* algorithm was used which adapts itself to differences in fixation detection thresholds on a subject-by-subject basis.

## 3 Results

### 3.1 Fixation Duration Distribution Analysis: Descriptives

To find possible differences in the distributions of fixation durations in eSports athletes with different skill level, we computed descriptive statistics for the individual fixation duration distributions. The averaged statistics for all three skill levels and the results of testing for significant differences (ANOVA) are presented in Table 1.

**Table 1.** Descriptive statistics for the individual fixation duration distributions averaged over skill levels

| Skill Level | Mean | Median | SD | Minimum | Maximum |
|---|---|---|---|---|---|
| Low | 250 | 213 | 122 | 117 | 800 |
| High | 278 | 223 | 197 | 72 | 1188 |
| Pro | 282 | 218 | 223 | 55 | 1229 |
| $F_{(2,18)}$ | 0.53 | 0.04 | 9.72 | 3.53 | 4.54 |
| p | n.s. | n.s. | <0.001 | 0.051 | <0.05 |

This analysis shows that mean and median durations are not different between skill levels. Differences are observed for the measures of the variability of duration. In the low skill group, minimum durations are higher and maximum durations are lower than in the more skilled groups. The professional group is characterized by very low minimal durations (technically, 50 ms is a typical cut-off for the identification of a fixation) and very large maximum durations. The standard deviations of the durations show a systematic increase with skill level. Thus, higher skill levels are characterized by more "wide" duration distributions with more variability.

Additional analysis using post-hoc multiple comparison test (Sheffe's test) showed that the differences in the standard deviations are differences between the low skill group and both the high skill and the pro group, that the differences in the minimums are mainly differences between the low skill group and the pro group, and that the differences in the maximums are again driven by the differences between the low skill group and both the high skill group and the pro group. This suggests that the pro and the high skill group are relatively similar while the low skill group is distinctively different from them. Overall, these results suggest that higher skill level leads to wider fixation duration distributions.

### 3.2 Vincentile Analysis

A more formal analysis of a frequency distribution form may be given by a Vincentile analysis [2]. The Vincentile analysis consists in separating distributions in five equal bins ("the vincentiles", each comprising 20% of the distribution), computing averages over the bins, and statistically analyzing differences in these averages. For example, if there are significant differences in the first vincentile, than the left tail of the distribution with a smaller first vincentile is shifted to the left.

We computed mean fixation durations over five bins (Figure 2). The previous analysis suggests that the line connecting bin means for the pro group should start lower than the line connecting the means for the bin means for the low skill group and cross it. The line for the high skill group should be in between these two lines. This is exactly what is seen in the Figure 2. We submitted the binned data to a two-way 3×5 rm-ANOVA and obtained a highly significant Bin × Skill Level interaction, $F(8,72)=10.3$, $p<0.001$. Generally, these results formally support the idea that progressing skill level makes the fixation duration distribution more wide.

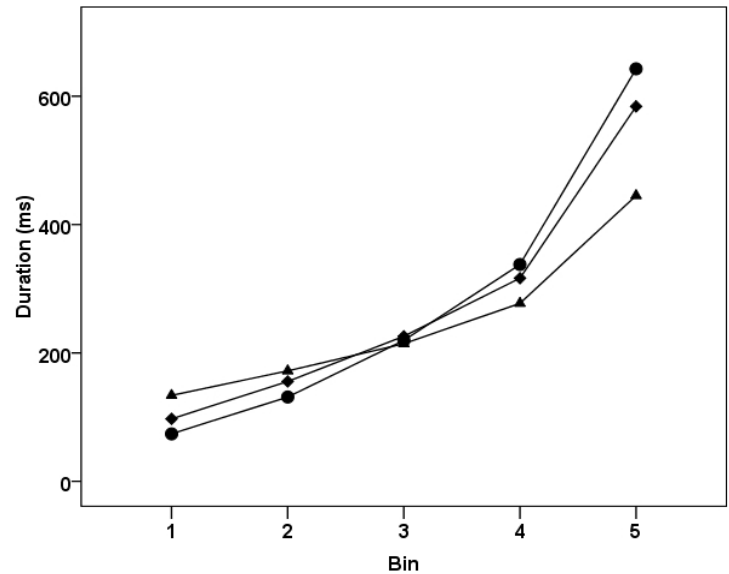

**Fig. 2.** Mean fixation durations (ms) for the five bins for different skill levels (low – triangles, high – diamonds, pro – circles)

### 3.3 Kernel Density Estimation Analysis

We further also applied a kernel density estimation procedure to the fixation duration distribution in the three skill groups. Kernel density estimation is a statistical procedure to compute a probability distribution of a random variable (like fixation duration) from a finite number of observations. Kernel density estimation is often used to reveal whether the "true" distribution of a variable is unimodal or bimodal. That is, it is used to assess whether the distribution of random variable values is homogenous or there are two (or more) clusters of values.

To analyze the specific form of fixation duration distributions at different skill levels we applied kernel density estimation to raw frequencies data. We were specifically interested in obtaining bimodal density plot for higher skilled groups as these may indicate the presence of two type of fixations – ambient (lower mean durations around 100 ms) and focal (higher mean durations around 300-350 ms). These two types of fixations have often been reported in eye-movements research. They seem to reflect either unconscious spatial processing (ambient) or conscious identification of visual obkects and events (focal). To this end, the standard *density()* function from the R statistical computing environment was used with the Gaussian kernel and the bandwidth estimated with the 'js' method (see Figure 3).

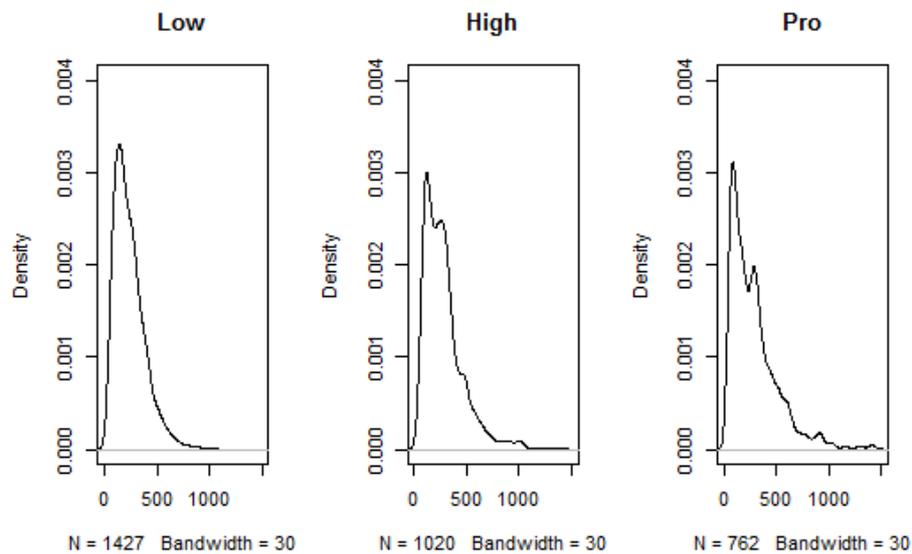

**Fig. 3.** Probability densities of fixation durations (in ms) in the low, high, and pro player groups estimated with the Gaussian kernel at bandwidth 30.

Overlaying the densities (not shown) reveals that they coincide to a large extent which explains why the above analysis of the means and medians didn't reveal significant differences. However, the form of distributions is clearly different suggesting the existence of two clusters of fixation durations in the higher skill groups. This is in line with the distinction between short ambient fixations and long focal fixation. This is also supported by the finding that the modes in the higher skill distributions are about 100 ms and 300 ms as is typically reported for ambient and focal fixations [12]. Overlaying the density of the low skill group over the pro group also indicates that there are more "very long" (over 500 ms) fixations in the pro group. Actually, these are virtually not present in the low skill group, contrary to the pro group.

## 4   Discussion

The results presented above suggest a clear difference in the distribution of fixation durations between low skilled amateurs, advanced amateurs, and professional eSports athletes while engaged in a realistic gaming experience. The low skill group shows a typical left shifted unimodal distribution of fixation durations. Fixation durations are mostly within the typical 200-400 ms range. With progressing skill level, there is a clear tendency for the variability in fixation durations to become larger. That is, there are more short fixations and more long fixations in the high skilled amateurs and, even more pronounced, in the professional gamers. A specific characteristics of the both advanced groups is the presence of very long fixations over 500 ms which are virtually absent in the low skill group.

A more detailed analysis of the distributions (kernel density estimation) revealed that the source of increased variability of fixation durations in the higher skilled groups may be the presence of different types of fixations in the latter. That is, in the low skill group the fixation duration distributions is highly uniform and unimodal. This suggests the presence of only one fixation cluster in this skill group. In the high skilled amateurs and, especially, in the professional players the distributions are bilmodal with the two modes around 100 and 300 ms. Thus, there may be two clusters of fixations typically associated with short (100 ms) and long (300 ms) fixations in the two higher skilled groups. This is a qualitative difference in the distribution of fixation duration which may depend on the skill level of cyberathlets which deserves further exploration.

There is a well-founded distinction of visual fixations into ambient fixations and focal fixations [3,5]. Ambient fixations are associated with spatial analysis of visual scenes, specifically, with the localization of objects. Ambient visual fixations are driven largely by automatic processes, are associated with the dorsal "Where" visual processing system in the human brain [6], and don't permit for conscious identification of objects. On the contrary, focal fixations reflect conscious perception of objects and are associated with the conscious ventral visual processing system in the brain (the "What" system). Studies have shown the importance of this distinction for predicting visual memory success and traffic accidents, for example [12]. Objectively, ambient fixations are short (clustering in the range 50-150 ms) and focal fixations are long (clustering around 250-350 ms). This allows identifying the two fixation clusters obtained in the both high skill groups as ambient and focal.

The ability to assess eSports athletes skill level based on the analysis of fixation duration distributions (and, generally, to assess the skill level in performing any complex computer-mediated visual interactions) may be of some importance for HCI. It may serve as input to developing personalized interfaces which adapt to the skill level of a user. Such interfaces may be adaptive to changes in the user's skill level which inevitably will occur over time. It is also important for supporting visual-motor interactions with computers as individually-tailored identification of ambient and, important, focal fixations will inform the interface what interface events were consciously processed by the user. Such analyses may also be of importance in interface design in order to assess the effectiveness of visual scanning of different interface layouts.

Game interface design solutions may be based on the analysis of the fixation durations. First, if an important screen element is mostly covered with short (ambient) fixation, it may be made visually more salient (size/color/animation) to grab players' focal attention. Second, a clear conjecture in interface design is that display elements which are focally fixated are subjectively the most important ones and so must be design with more care. Third, the ratio of ambient of focal fixations may reveal the cognitive load players experience when navigating and acting in the virtual world. It is to note that we choosed a highly specific shooter game (CS:GO) as a case study.

So, our results may not generalize directly to other kinds of video games. However, in any game which employs sophisticated visualizations, there will be room for a detailed analysis of players' eye movements in order to improve the interface.

## 5    Conclusions

eSports athletes with different skill levels (low, high, and professional) play computer game under realistic conditions. Eye movements were recorded and the distributions of fixation durations were analyzed. It was found that increasing skill leads to more variable fixations durations (more shorter and more longer fixations). It was also found that high skill was characterized by bimodal fixations durations distributions suggesting the presence of two fixation types. These fixation types were identified as ambient and focal fixations. The results were interpreted as indicating a complex interplay of automatic ambient visual processing and conscious focal visual processing in high skilled gamers. The objective identification of skill level of users performing complex visual-motor interactions with computers may be used in developing adaptive user interfaces and in interface design.

## 6    Acknowledgements

The reported study was funded by RFBR according to the research project No. 18-29-22077\18. Authors would like to thank Skoltech Cyberacademy, CS:GO Monolith team and their coach Rustam ``TsaGa'' Tsagolov for fruitful discussions while preparing the paper.The authors thank Alexey "ub1que" Polivanov for supporting the experiments by providing a slot at the CS:GO Online Retake server.к.

## 7    References

1. Al-Samarraie, H., Sarsam S.M., & Guesgen, H. (2016). Predicting user preferences of environment design: a perceptual mechanism of user interface customization. Behaviour & Information Technology, 35:8, 644-653, DOI: 10.1080/0144929X.2016.1186735
2. Balota, D.A., Yap, M.J., Cortese, M.J. & Watson, J.M. (2008). Beyond mean response latency: Response time distributional analyses of semantic priming. J Mem Lang, 59, 495-523.
3. Berman, R. & Colby, C. (2009). Attention and Active Vision. Vision Research, 49(10), 1233–1248.
4. Eisenberg, M.L, & Zacks, J.M. (2016). Ambient and focal visual processing of naturalistic activity. Journal of Vision, 16(2):5. doi: 10.1167/16.2.5.
5. Fitts, P. M., & Posner, M. I. (1967). Human performance. Belmont, CA: Brooks/Cole.
6. Follet, B., Le Meur, O., & Baccino, T. New insights into ambient and focal visual fixations using an automatic classification algorithm. Iperception. 2011; 2(6): 592–610. doi: 10.1068/i0414
7. Goodale, M.A, & Milner, A.D. (1992). Separate visual pathways for perception and action. Trends in Neurosciences, 15(1), 20–25.

8. Graham, T. C. N., Curzon, P., Doherty, G., Palanque, Ph., Potter, R., Roast, C., & Smith, S. P. (2007). Usability and Computer Games: Working Group Report. In: DSV-IS 2006: Interactive Systems. Design, Specification, and Verification. Lecture Notes in Computer Science, Vol. 4323. Pp. 265-268. Heidelberg: Springer.
9. Jacob, R.J.K. (1991). The Use of Eye Movements in Human-Computer Interaction Techniques: What You Look At is What You Get. ACM Transactions on Information Systems, 9(3), 152-169.
10. Roccetti, M., Marfia, G., & Semeraro, A. (2012). Playing into the wild: A gesture-based interface for gaming in public spaces. Journal of Visual Communication and Image Representation, 23(3), 426-440.
11. Starke, S.D., & Baber, C. (2018). The effect of four user interface concepts on visual scan pattern similarity and information foraging in a complex decision making task. Applied Ergonomics, 70, 6-17.
12. Van Renswoude, D.R., Raijmakers, M.R.F., Koornneef, A., Johnson, S. P., Hunnius, S., & Visser, I. Gazepath: An eye-tracking analysis tool that accounts for individual differences and data quality. Behavior Research Methods, 50(2): 834–852.
13. Velichkovsky, B.M., Dornhoefer, S.M., Kopf, M., Helmert, J., & Joos, M. Change detection and occlusion modes in road-traffic scenarios. Transportation Research Part F: Traffic Psychology and Behaviour, 5:2, 99-109.
14. Velichkovsky, B.M., Dornhoefer, S.M., Pannasch, S., & Unema, P. (2000). Visual fixations and level of attentional processing. Proceedings of the Eye Tracking Research & Application Symposium, ETRA 2000, Palm Beach Gardens, Florida, USA, November 6-8, 2000. DOI: 10.1145/355017.355029